\documentclass{emulateapj} 
\usepackage{psfig}
\usepackage{bm} 
\usepackage{aalongtable}
\usepackage{rotating}
\usepackage{times}
\usepackage{amsmath,amssymb}
\usepackage{color}
\usepackage[colorlinks=true, linkcolor=blue, citecolor=blue, urlcolor=blue]{hyperref}

\def\apj{ApJ}% % Astrophysical Journal
\def\apjl{ApJL}% % Astrophysical Journal, Letters
\def\apjs{ApJS}% % Astrophysical Journal, Supplement
\def\ao{Appl.~Opt.}% % Applied Optics
\def\aap{A\&A}% % Astronomy and Astrophysics
\def\mnras{MNRAS}% % Monthly Notices of the RAS
\def\pasp{PASP}% % Publications of the ASP
\DeclareMathAlphabet{\mathsc}{OT1}{cmr}{m}{sc}
\def\testbx{bx}%
\DeclareRobustCommand{\ion}[2]{%
\relax\ifmmode
\ifx\testbx\f@series
{\mathbf{#1\,\mathsc{#2}}}\else
{\mathrm{#1\,\mathsc{#2}}}\fi
\else\textup{#1\,{\mdseries\textsc{#2}}}%
\fi}

\newcommand{\hb} {\mbox{H$\beta$}}

\newcommand{\Feii} {\ion{Fe}{ii}}

\newcommand{\Nai} {\ion{Na}{i}}

\newcommand{\Mgii} {\ion{Mg}{ii}}
\newcommand{\Civ} {\ion{C}{iv}}

\newcommand{\Oiiia} {[\ion{O}{iii}]}

\newcommand{\ebv}{\text{$E(B-V)$}}

\newcommand{\msun}{\text{M$_{\odot}$}}

\newcommand{\kms}{\text{$\rm{\,km\,s^{-1}}$}}

\shorttitle{A double double radio quasar} 
\shortauthors{Nandi et al.}

\begin{document}


\title{Discovery of a red quasar with recurrent activity}

% from AASTeX v4.0. You can use \email to mark an email address 

\author{
S. Nandi$^\star$\altaffilmark{1,2,3},
R. Roy\altaffilmark{4},
D.J. Saikia\altaffilmark{5,6},
M. Singh\altaffilmark{2},
H.C. Chandola\altaffilmark{3},
M. Baes\altaffilmark{1},
R. Joshi\altaffilmark{2}, 
G. Gentile\altaffilmark{1,7} and 
M. Patgiri\altaffilmark{8}
}

\email{$^\star$sumana1981@gmail.com, Sumana.Nandi@UGent.be}

\altaffiltext{1}{Sterrenkundig Observatorium, Universiteit Gent, Krijgslaan 281 S9, B-9000 Gent, Belgium}

\altaffiltext{2}{Aryabhatta Research Institute of Observational Sciences (ARIES), Manora Peak, Nainital, 263 002, India}

\altaffiltext{3}{Department of Physics, Kumaun University, Nainital 263 001, India}

\altaffiltext{4}{Institut d'Astrophysique et de G\'{e}ophysique, Universit\'{e} de Li\`{e}ge, All\'{e}e du 6 Ao\^{u}t 17, B\^{a}t B5c, 4000 Li\`{e}ge, Belgium}

\altaffiltext{5}{National Centre for Radio Astrophysics, TIFR, Pune University Campus, Post Bag 3, Pune 411 007, India}

\altaffiltext{6}{Cotton College State University, Panbazar, Guwahati 781 001, India}

\altaffiltext{7}{Department of Physics and Astrophysics, Vrije Universiteit Brussel, Pleinlaan 2, 1050 Brussels, Belgium}

\altaffiltext{8}{Cotton College, Panbazar, Guwahati 781 001, India}

% abstract
%___________________________________________________________________________

\begin{abstract}
 We report a new double-double radio quasar, DDRQ, J0746$+$4526 which exhibits
 two cycles of episodic activity. From radio continuum observations at
 607 MHz using the GMRT and 1400 MHz from the FIRST survey we confirm
 its episodic nature. We examine the SDSS optical spectrum and estimate the 
 black hole mass to be (8.2$\pm$0.3)$\times$10$^7$M$_\odot$ from its observed
 MgII emission line, and the Eddington ratio to be 0.03. The black hole mass
 is significantly smaller than for the other reported DDRQ, J0935+0204, while
 the Eddington ratios are comparable. The SDSS spectrum is significantly red
 continuum dominated suggesting that it is highly obscured with
 ${\ebv}_{host}=0.70\pm0.16$ mag. This high obscuration further indicates the
 existence of a large quantity of dust and gas along the line of sight, which
 may have a key role in triggering the recurrent jet activity in such objects.
\end{abstract}	




% sec:intro
%___________________________________________________________________________
\section{introduction} \label{sec:intro}
 
 It is well established that many active galactic nuclei (AGN) may go through
 two or more cycles of episodic activity
 \citep{1996MNRAS.279..257S, 1999A&A...348..699L}. This is most clearly seen in
 extended radio galaxies and quasars, where there may be a new pair of radio
 lobes with well-defined hot-spots closer to the nucleus, in addition to the
 more distant and diffuse, extended lobes from an earlier cycle of activity.
 The sources with a second pair of lobes have been classified as double-double
 radio galaxies (DDRGs) by \citet{2000MNRAS.315..371S}, and their properties
 have been summarized by \citet{2009BASI...37...63S}. Although most DDRGs were
 believed initially to be giant radio sources with sizes of over about a Mpc,
 a significant number of smaller sized DDRGs have also been identified in recent
 years \citep{2012BASI...40..121N}. In most of the cases double-double radio
 sources are associated with galaxies but it is also possible for a quasar to
 appear as a radio-source exhibiting episodic activity. However, the number of
 DDRQs reported so far is very limited \citep{2009MNRAS.399L.141J}, and it is
 important to identify more of these to make the physical scenario of this class
 statistically robust. J0935$+$0204 (4C02.27) located at the redshift of 0.65, is one promising
 DDRQ reported by \citet{2009MNRAS.399L.141J}, has a blue-continuum
 dominated spectrum, with projected linear sizes of 70 and 470 kpc for the inner
 and outer radio lobes respectively. Although in optical surveys, the widely
 popular ``Color selection technique'' already identified a large number of
 quasars having blue-continuum dominated spectra, recent studies found a
 huge dust accumulation around many new similar objects which make them
 optically redder \citep{2009ApJ...698.1095U}. These red quasars are usually 
 believed to represent an early stage of AGN evolution, which is going through 
 a merging process leading to presence of large amounts of gas and dust \citep{2012ApJ...757...51G}. 

 Based on only radio structural information \citet{2011ApJS..194...31P}
 classified 811,117 radio sources or entries from the FIRST (Faint Images of the
 Radio Sky at Twenty-cm; \citealt{1995ApJ...450..559B}) survey into different
 categories and listed 242 sources as candidate DDRGs. Further detailed
 investigation of these sources along with optical data from
 SDSS\footnote{http://www.sdss3.org/dr9/} and
 DSS\footnote{http://archive.eso.org/dss/dss} catalogues showed only 23 of these
 sources to be promising examples of DDRGs \citep{2012BASI...40..121N}. In this
 paper we report the discovery of a red DDRQ J0746+4526, discuss some of its
 properties and also compare these with J0935$+$0204. J0746$+$4526 which was
 classified from this survey to show evidence of episodic activity, has been
 identified as a quasar at a redshift 0.55021$\pm$0.00005 in Data Release 9 of
 SDSS. However, to confirm that neither of the inner compact components is a
 compact flat-spectrum nuclear component, as appeared to be the case in many of
 the candidate DDRGs identified by \citet{2011ApJS..194...31P}, we observed this
 source with the Giant Metrewave Radio Telescope (GMRT) at 607 MHz. 

 We present the results of these observations of J0746$+$4526, which confirm the
 source to be a DDRQ, and then discuss the nature of the source using the
 results of our observations and archival FIRST data at radio wavelengths, and
 optical data from SDSS. The observations and radio data reduction procedures
 are described in \S\ref{sec:obs_data}. In \S\ref{sec:obs_rsult}, we present the
 observational results. The black hole mass determination is described in
 \S\ref{sec:Mbh}. The discussion and concluding remarks are summarized in
 \S\ref{sec:discussion}.

%--------------------------------------------------------------------------
%____________________________________________________________________________
\begin{figure*}
\centering
\vbox{
\hbox{
\includegraphics[width=2.75in,angle=0]{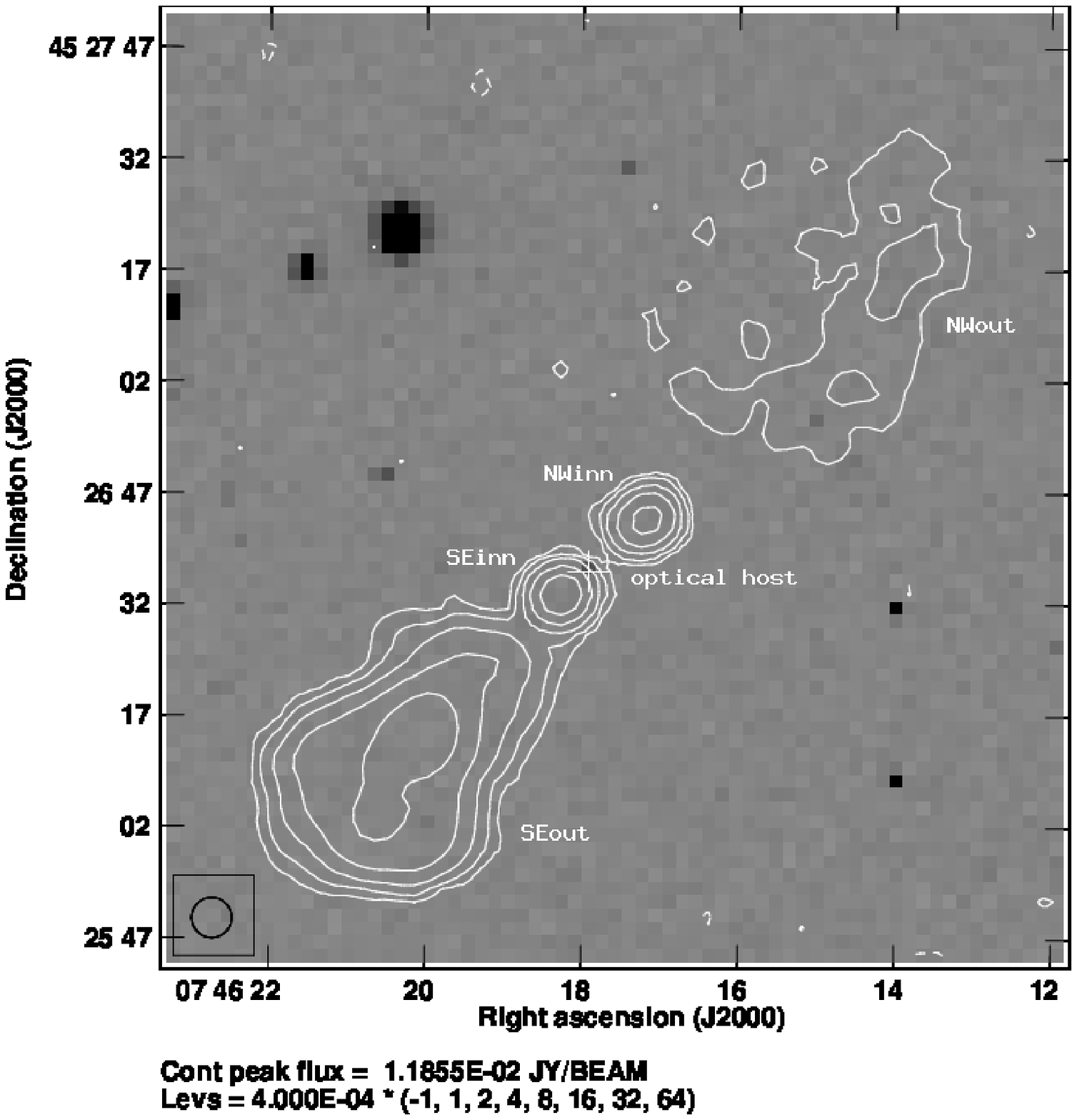}
\hskip 10mm
\includegraphics[width=2.8in,angle=0]{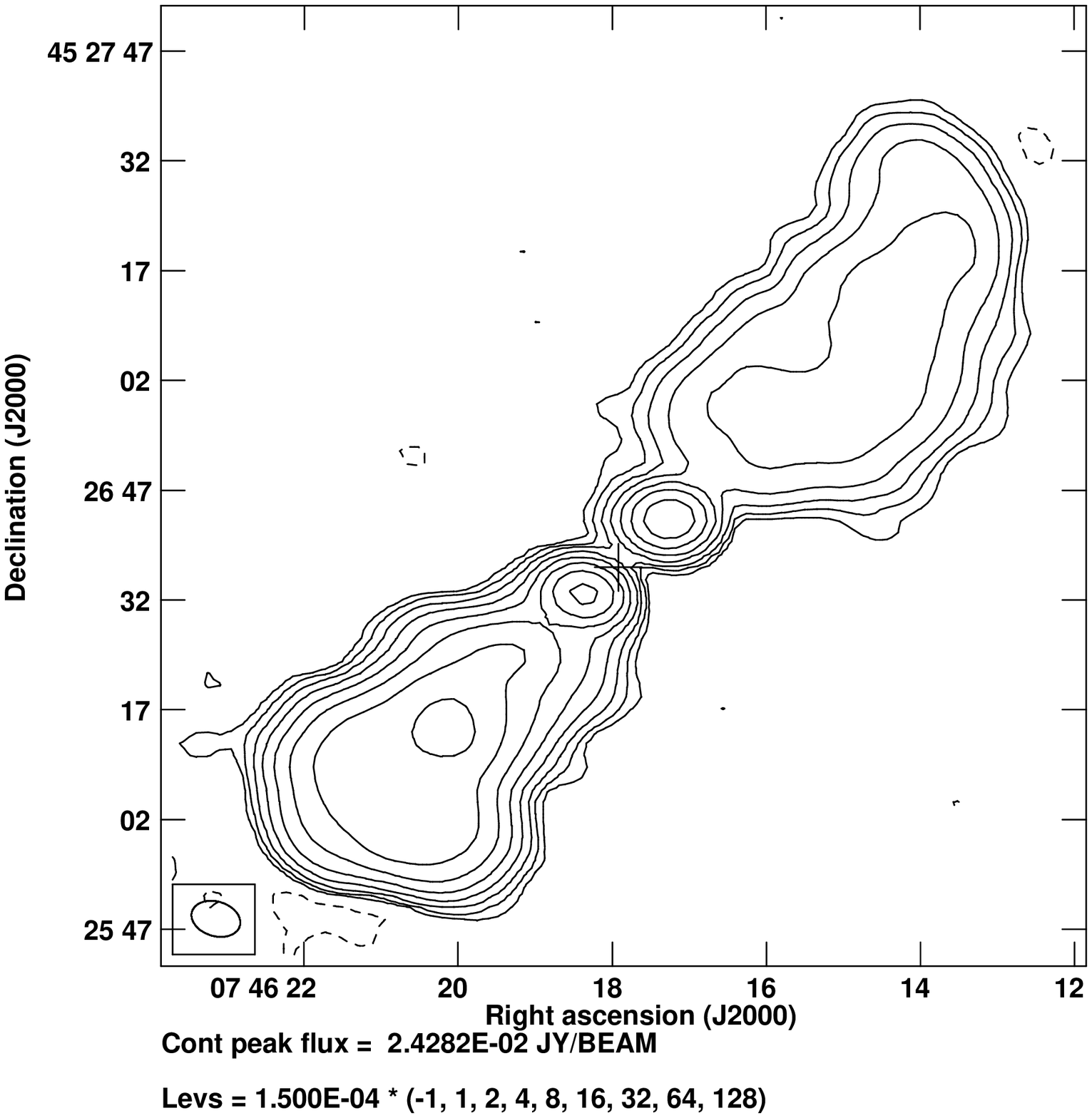}
     }
     }
\vskip 5mm
\caption[]{Left panel: The FIRST image at 1400 MHz overlayed on the optical
 field from the SDSS. The optical host and both the outer and inner doubles
 (NW$_{out}$, SE$_{out}$, NW$_{inn}$, SE$_{inn}$) are marked in the same image.
 Right panel: The GMRT image at 607 MHz. The $+$ sign represents the host
 position.} 
\label{fig:FIRST_GMRT}
\end{figure*}
% __________________________________________________________________________
%--------------------------------------------------------------------------

\section{Observations and data reduction} \label{sec:obs_data}
 We made the 607 MHz GMRT observations of J0746$+$4526 on 2012 Nov 29. The
 observations were made in the standard manner, with each observation of the
 target source interspersed with observations of the phase calibrator,
 J0713+438. After each of several 20 min exposures of the target source the
 phase calibrator was observed for 5 min. The total observing time on the target
 source is about 190 min. We observed 3C147 for flux density as well as bandpass
 calibration. All flux densities are on the \citet{1977A&A....61...99B} scale
 using the latest VLA values. These data were edited and reduced with the NRAO
 AIPS package. Several rounds of self calibration were done to produce the best
 possible images.

%--------------------------------------------------------------------------
\begin{table}
\caption{The observational parameters obtained from radio data$^\dagger$.}
\label{tab:obsinfo}
\begin{tabular}{llllllcc}
\hline
Freq.& \multicolumn{2}{c}{Beam size} & PA & rms&Cmp.&S$_p$&S$_t$\\
(MHz)&($^{\prime\prime}$)&($^{\prime\prime}$)&($^\circ$)&(mJy/b)&&(mJy/b)&(mJy)\\
(1)&(2)&(3)&(4)&(5)&(6)&(7)&(8)\\
\tableline
607  &6.83&4.71&73&0.05 & NW$_{\text{out}}$ & 3& 93\\
     &    &    &  &    & NW$_{\text{in}}$ & 17& 24\\
     &    &    &  &    & SE$_{\text{in}}$ & 24& 30\\
     &    &    &  &    & SE$_{\text{out}}$ & 24&319\\
1400 &5.40&5.40& 0&0.14 & NW$_{\text{out}}$ & 1& 24\\
     &    &    &  &    & NW$_{\text{in}}$ & 8& 11\\
     &    &    &  &    & SE$_{\text{in}}$ &12& 13\\
     &    &    &  &    & SE$_{\text{out}}$ &10&119\\
\tableline
\end{tabular}\\\\
$^\dagger${Arrangement of the Table is as follows. Column 1: frequencies of
 observations; columns 2$-$3: the angular sizes of major and minor axes of the
 restoring beam; column 4: The Position angle; column 5: the rms noise; column
 6: component designation; column 7: the peak flux density; column 8: the total
 flux density of each component. The error in the flux density is approximately
 7\% at 607 MHz and 5\% at 1400 MHz.}

\end{table}
%--------------------------------------------------------------------------

\section{Observational results}
\label{sec:obs_rsult}

\subsection{Radio data}
\label{sec:rdodata}
 
 The full-resolution radio image obtained from FIRST at 1400 MHz and that from
 GMRT at 607 MHz are presented in Figure \ref{fig:FIRST_GMRT}. The diffuse
 north-western outer lobe is imaged better at the lower-frequency GMRT image.
 The observational parameters and the flux densities estimated from these
 images are presented in Table \ref{tab:obsinfo}. The spectral indices
 $\alpha$ (defined as $S_\nu\propto\nu^{-\alpha}$) between 607 and 1400 MHz for
 the NW$_{\text{out}}$ (north-west outer) and SE$_{\text{out}}$ (south-east
 outer) lobes are 1.62$\pm0.13$ and 1.18$\pm0.13$ respectively. The
 corresponding two-point spectral indices for the NW$_{\text{inn}}$ (north-west
 inner) and SE$_{\text{inn}}$ (south-east inner) components are 0.93$\pm$0.13
 and 1.00$\pm$0.13 respectively.
  
 The outer-doubles of this DDRQ appear reasonably well aligned with the
 inner-ones and collinear with the parent optical host galaxy. The radio core
 has not been detected at either frequency. The projected linear size between
 the outer radio lobes is $\sim$630 kpc whereas the separation between two inner
 lobes is $\sim$95 kpc \footnote{We have assumed a Universe with H$_0$=71
 km s$^{-1}$ Mpc$^{-1}$, $\Omega_{\rm m}$=0.27 and $\Omega_{\rm vac}$=0.73}.
 The outer lobes are highly asymmetric in intensity and do not show any evidence
 of hot-spots at the ends of the lobes. The flux density ratio of the outer
 components at 1400 MHz is 4.96, while for the inner lobes it is 1.18, with the
 south-eastern component being brighter in both cases. The log luminosity at an
 emitted frequency of 1400 MHz for the inner and outer doubles are 25.41 and
 26.17 W/Hz respectively \citep{2012BASI...40..121N}.
 
% The variation of spectral index over the entire radio source (Figure
% \ref{fig:dist_spectralindx}) has been estimated by
% convolving both images to a uniform resolution of 7$\arcsec$, and splitting the
% extended lobes into strips, each with a width of 7$\arcsec$. For the inner
% double, a single strip has been used for each component. The variation of
% spectral index appears complex, with the north-western lobe being steeper, but
% exhibiting evidence of flattening towards the outer edges. The south-eastern
% lobe, on the other hand, shows no significant variation within $\sim$250 kpc
% from the host galaxy, but steepens rapidly beyond this distance. A more
% detailed spectral index study over a large frequency range would be desirable
% to understand and model the variations.

\subsection{Optical data}
\label{sec:optdata}

 The optical analysis is based on archival spectra obtained from the SDSS DR9
 catalog for both the DDRQs J0746+4526 and J0935+0204. The Galactic reddening
 $E(B-V)$, adopted from the NASA Extragalactic Database (NED), towards
 J0746+4526 and J0935+0204 are respectively 0.056 and 0.042 mag. 

\begin{figure}
\centering
\includegraphics[width=7.0cm,angle=0]{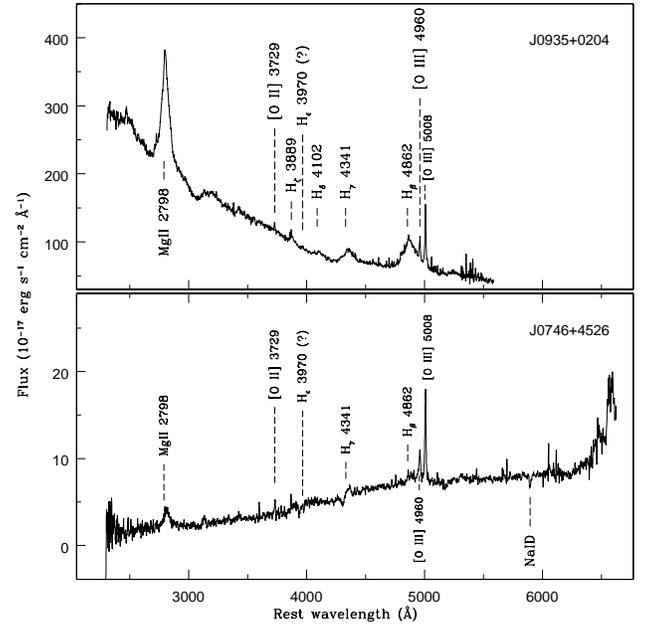}
\caption{The SDSS spectra of J0935+0204 (upper panel) and J0746+4526 (lower
 panel). For both spectra, the wavelengths have been corrected for the redshifts
 of the galaxies, while the fluxes have been mentioned after performing the
 corrections due to corresponding redshifts and galactic extinctions. Several
 spectral lines have been marked in both spectra.}
\label{spec_com}
\end{figure}

 The Galactic extinction and redshift corrected optical spectrum of J0746+4526
 is presented in Figure \ref{spec_com}, along with the spectrum of J0935+0204.
 The spectrum of J0935+0204 is similar to the typical spectra of normal quasars,
 while the spectrum of J0746+4526 is typical of red quasars. For both quasars,
 the \Oiiia\ lines at  $\lambda$4960\AA\ and $\lambda$5008\AA\ associated with
 the narrow-line-region (NLR) clouds, are quite prominent, while there is a
 striking difference in the strength of the hydrogen Balmer lines which are
 weaker in J0746+4526. It should be noted that the spectra of both the quasars
 are of moderate signal to noise ratio (S/N). For J0746+4526 and J0935+0204 it
 is $\sim$13.6 and $\sim$26.5 respectively, measured around mid portion of the
 SDSS spectral range coverage. 

 We have estimated the reddening
 for the object J0746+4526 in two different ways. First we use the empirical
 relation, developed from the study of spectra of nearby supernovae
 \citep[e.g.][]{1990A&A...237...79B, 2003fthp.conf..200T, 2011MNRAS.414..167R, 2012MNRAS.426.1465P}, between the width of the \Nai\,D absorption dip and the
 host reddening. This has been used widely to determine the host galaxy
 extinction directly from the observed spectrum. The presence of the \Nai\,D
 lines ($\lambda\lambda$5890, 5896\AA) in the spectrum of J0746+4526, for which
 we obtain an equivalent width of 3.45 \AA, enables us to estimate the plausible
 host extinction directly from the observed spectrum. Using the relation
 prescribed by \citet{2003fthp.conf..200T}, this corresponds to a host reddening
 of approximately 0.54 mag. On the other hand, according to the relation
 proposed by \citet{1990A&A...237...79B}, the reddening of the host is
 $\sim 0.86$ mag. We therefore assume that the mean host reddening along the
 line of sight is ${\ebv}_{\text{host}}=0.70\pm0.16$ mag. 

 We have alternatively estimated the amount of foreground extinction by
 comparing the de-reddened spectrum of J0746+4526 with an appropriate template
 (composite spectrum obtained from
 \citet{2001ApJ...546..775B}\footnote{http ://sundog.stsci.edu/Ðrst/QSOComposites/.}). To do so we have adopted the Small Magellanic Cloud extinction curve, and
 we have de-reddened our observed spectrum by considering dust distributed in
 front of the QSO in a uniform screen geometry. We have searched for the best
 value of the color excess that minimizes the differences between the template
 and our QSO spectrum. We find that the host reddening is $\sim$ 0.79 mag. This
 result is consistent with the result obtained from the \Nai\,D absorption dip.
 This is also consistent with the typical reddening ($\sim 0.1-1.5$ mag)
 determined for the red quasars \citep{2009ApJ...698.1095U}.

 The widths of several spectral lines which are presumed to be generated near
 the central region of the AGN (like \Mgii\,$\lambda$2800\AA\ and H Balmer
 lines) are widely different for two interesting objects. For J0746+4526, FWHM
 of \Mgii\,$\lambda$2800\AA\ is $\sim$ 35.22\AA\,, whereas for J0935+0204 it is
 about 61.33\AA\ \citep{2012MNRAS.422.1546K}. These demonstrate that the
 velocity dispersions of several broad-line-region (BLR) clouds are much higher
 for J0935$+$0204 in comparison to that for J0746+4526 and further indicate that
 the former galaxy hosts a higher mass central black hole than the latter one. 
%--------------------------------------------------------------------------
%____________________________________________________________________________
%\begin{figure}
%\centering
%\includegraphics[width=4.5cm,angle=-90]{./Fig.2a.eps}
%\vskip 4mm
%\includegraphics[width=5.35cm,angle=-90]{./Fig.2b.eps}
%\vskip 5mm
%\caption{The variation of spectral index over radio-lobes of J0746+4526. The
% Upper panel presents the spectral index map produced by using 607 MHz and 1400
% MHz images. The outer double appears to have a relatively steeper spectrum than
% the inner double. The Lower panel presents the spectral indices in the
% consecutive strips along the lobes vs. distance from the optical host. The
% filled circles represent the outer doubles, while the open circles represent
% the inner doubles. The variation in spectral index are from 0.5 to 2.0.}
%\label{fig:dist_spectralindx}
%\end{figure}
%____________________________________________________________________________
%--------------------------------------------------------------------------

%--------------------------------------------------------------------------
%------------------------------------------------------------------------------%

\section{Estimation of the Black Hole mass and Eddington ratio} 
\label{sec:Mbh}
 ``Reverberation Mapping'' (RM) or ``Echo Mapping'' has proven to be a
 viable technique to measure the location of the line emitting clouds and the
 black hole mass \citep{1982ApJ...255..419B, 1993PASP..105..247P}. A major
 drawback with RM is that for a single system it requires a long-term monitoring
 programme to measure the time gap between continuum and the broad line
 variability. To measure the black hole mass of this DDRQ we follow an indirect
 technique of virial single-epoch method which is an approximation to the
 reverberation mapping method
 \citep{ 2005ApJ...629...61K,2010MNRAS.402.1059C, 2011MNRAS.412.2717J,
 2012MNRAS.426..851K}. It exploits the empirical
 power-law correlation between the size of the BLR and the AGN continuum
 luminosity (R$_{\text{BLR}}$ $\propto$ $L^{\beta}$; with $\beta \sim$ 0.5), as expected
 from photoionization model predictions \citep{2006ApJ...644..133B}.

 The best calibration values for the single epoch mass measurement scaling
 relations are available for the \hb\ line, as it is the basis for the majority
 of reverberation mapping programmes. However, the similar R$_{\text{BLR}}$-L relationship
 is also frequently employed to measure the BH mass using the \Civ\ and \Mgii\
 lines. The redshifts of these DDRQs are such that only the \Mgii\
 $\lambda\lambda2798,2803$\AA\AA\ and \hb\ $\lambda4861$\AA\ lines fall in the
 wavelength range of the observed spectra. We note that the \hb\ and \Mgii\
 emission features are prominent in the spectrum of J0935+0204, whereas for
 J0746+4526 the \Mgii\ line is clearly seen (although weaker) and the \hb\
 emission line is barely seen (see Figure  \ref{spec_com}). Therefore, to
 determine the black hole mass of J0746+4526, we have used the \Mgii\ emission
 line, whereas for J0935+0204,
 \citet{2012MNRAS.426..851K} used both the \Mgii\ and \hb\ emission lines.

 According to the \citet{2009ApJ...699..800V} (followed by
 \citealt{2012MNRAS.426..851K}), the FWHM of the \Mgii\ emission line of a
 quasar spectrum is related to its black hole mass by the following scaling
 formula :
\begin{equation}
\frac{M_{\text{BH}}}{\msun}
=
7.24 \times 10^6
\left[\frac{\lambda L_\lambda\,(3000\,\text{\AA})}{10^{44}~\text{erg}\,\text{s}^{-1}}\right]^{0.5} 
\left[\frac{{\text{FWHM}}_{\text{\Mgii}}}{1000~{\text{km}}\,{\text{s}}^{-1}}\right]^2
\label{eq1}
\end{equation}
where $L_\lambda\,(3000\,{\text{\AA}})$ is the monochromatic luminosity at 3000\,\AA\,.

 The \Mgii\ feature of J0746+4526 spectrum has been fitted with multiple
 Gaussian functions along with a power-law continuum and underlying UV \Feii\
 features have been obtained from the templates of nearby QSOs
 \citep{2006ApJ...650...57T}. In Gaussian profile fits, our initial guess
 consists of two Gaussian components for each line of the \Mgii\ doublet.
 However, if one of the components becomes statistically insignificant the
 procedure automatically drops it during the fit \citep{2010MNRAS.402.1059C}. In
 our two component Gaussian profile fits, width of each component (narrow/broad)
 of \Mgii\ 2796\AA\ were tied to the respective components of the MgII 2803\AA\
 line. Here broad and narrow components represent the \Mgii\ emission
 associated with BLR and NLR clouds respectively. The width of the broad
 components are set to be greater than 2000 \kms\, whereas those for the narrow
 components are constrained to be less than 1000 \kms\,. In addition, we have
 constrained the width of the broader \Mgii\ component to be the same as the
 width of UV \Feii\ emission line in this region. 

\begin{figure}
\centering
\includegraphics[width=7.0cm]{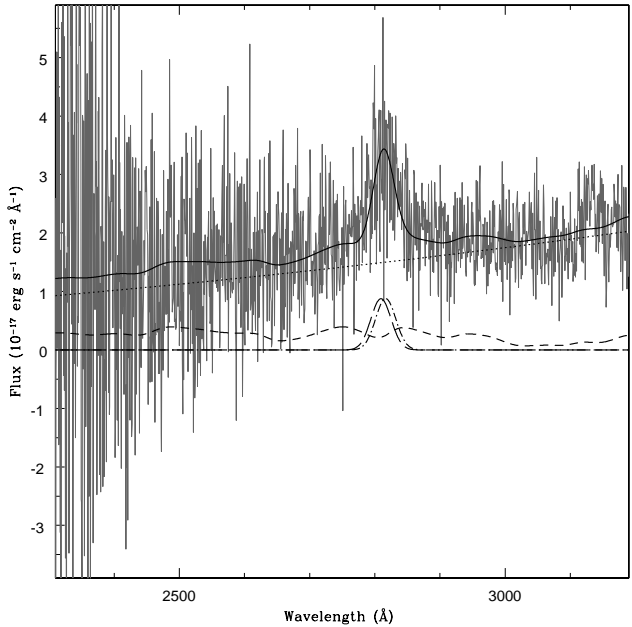}
\caption{The best fit to the \Mgii\ emission line of the SDSS spectrum of
 J0746+4526. The observed spectrum is presented in gray colour.
 The continuum fit is presented by dotted line. The long-dash and dot-dash lines
 represent the two broad emission features and the dash line marks the
 underlying \Feii\ emission. The modeled spectrum is presented as a continuous
 line.}
\label{J0746_MgII}
\end{figure}

 The fitting has been done in an iterative way, minimizing the reduced
 $\chi^2$ to $\approx 1$. Our final spectral fitting in the UV region is shown
 in Figure \ref{J0746_MgII}. Absence of the narrow \Mgii\ features in this
 fitting process probably indicates that no (or very negligible) \Mgii\
 emissions are produced by the NLR clouds associated with J0746+4526. The
 contribution of the underlying \Feii\ features is also very small. The FWHM of
 the broad components, obtained from the fitting, is 35.22$\pm$0.1 \AA\,. This
 corresponds to a BH mass of $(8.2\pm0.3)\times10^7$ \msun\,, where the error
 has been estimated after propagating the errors associated with flux estimation
 at 3000 \AA\ and FWHM estimation of \Mgii\ line in a quadrature. This value is
 about one order smaller than the BH mass estimated for J0935$+$0204
 ($(13.22\pm0.75)\times10^8$ \msun\ from \Mgii\ fit;
 \citealt{2012MNRAS.426..851K}).

 The Eddington ratio has been computed from the relation  $\ell = L_{\text{bol}}/L_{\text{E}}$, where
 $L_{\text{bol}}$ is the bolometric luminosity and $L_{\text{E}}$ is the
 Eddington luminosity \citep{2006MNRAS.365..101M, 2009ApJ...696.1998D, 2012MNRAS.426..851K}. 
 For J0746$+$4526, the computed values of $L_{\text{bol}}$ and $L_{\text{E}}$ are respectively $\sim 3.7\times10^{44}
 \text{erg}~\text{s}^{-1}$ and $\sim 1.2\times10^{46} \text{erg}~\text{s}^{-1}$. This implies for J0746$+$4526, the 
 Eddington ratio $\sim$ 0.03, which is consistent with that of J0935$+$0204 ($\sim$ 0.06 calculated from \Mgii\
 profiles; \citealt{2012MNRAS.426..851K}). 
   
%--------------------------------------------------------------------------
%------------------------------------------------------------------------------%

\section {Discussion and conclusions} 
\label{sec:discussion}
 In this paper we have presented a radio and optical spectroscopic study of
 the DDRQ J0746$+$4526. Both GMRT and FIRST images of
 this object show that there have been two episodes of activity. The
 diffuse emission of the outer double is well beyond both the hotspots of 
 the inner lobes. The steeper spectral indices of the outer lobes 
% (see Lower panel of Figure \ref{fig:dist_spectralindx}),
 suggest that these are due to older emission. The lobes of the outer double are
 significantly more asymmetric in flux density than the inner ones. 
 The core emission is not detected at either 1400 or 607 MHz. This is in 
 contrast with the DDRQ J0935$+$0204, where a prominent core has 
 been detected at both frequencies. Although higher resolution observations
 at higher frequencies would be required to estimate the core flux density
 reliably, non-detection of a core in J0746$+$4526 is consistent with a 
 larger angle of inclination of this source to the line of sight \citep{1994MNRAS.270..897S}.

 Although the projected linear size of J0746$+$4526 is larger than that of
 J0935$+$0204, both are of FRII type, lobe dominated DDRQs and highly
 asymmetric in flux density. J0935$+$0204 has a bright hotspot in the
 south-western outer lobe. This feature indicates that, for this source, the
 time scale of episodic activity is less than a few Myr
 \citep{2009BASI...37...63S, 2009MNRAS.399L.141J}. On the other hand, both the
 outer lobes of J0746$+$4526 do not show any hot-spots, and a more detailed
 multi-frequency study would be required to estimate its ages.
 
 J0746+4526 and J0935+0204 are two quasars where recurrent jet activity has been
 observed. The optical spectrum of J0935+0204 shows a blue continuum dominated
 spectrum, similar to most quasars, whereas that of J0746+4526 shows evidence of
 more extinction and obscuration, similar to that of red quasars. 
 Significant populations of obscured red quasars have been reported in recent studies 
 \citep[e.g.,][]{2009ApJ...698.1095U, 2012ApJ...757...51G}. These findings demonstrate
 that possibly some different mechanism rather than the orientation of the
 torus is responsible for large obscuration. In particular, the large extinction 
 in red quasars is probably related to significant dust extinction in the host 
 galaxies. The fact that, out of the two known DDRQs, one is a red quasar is an 
 interesting result. 
 As large dust extinction is often related to massive starbursts and galaxy merging, 
 it suggests that galaxy merging cannot only trigger powerful AGN activity 
 \citep{1986ApJ...311..526H, 1995ApJ...438...62W, 2002A&A...396..773S} 
 but that it can also trigger the interruption and restarting of jet formation.
 In other words, this result plausibly supports the merger-driven scenario for DDRG 
 formation \citep{2003MNRAS.340..411L}. In order to test whether this mechanism is the 
 main scenario for DDRG formation, a larger sample of DDRGs needs to be investigated and 
 compared with the global AGN population. The present paper, which presents the first 
 discovery of a red quasar with recurrent jet activity, is only the first step towards a 
 larger investigation.

 The BH masses of giant and small-sized radio quasars, measured from \Mgii\
 emission lines typically belong to the range
 $1.6\times10^8~\msun<M_{\text{BH}}<12.2\times10^8~\msun$ and 
 $1.0\times10^8~\msun<M_{\text{BH}}<20.3\times10^8~\msun$ respectively
 \citep{2012MNRAS.426..851K}. For J0746+4526, the BH mass estimated using the
 \Mgii\ line is smaller than that of J0935+0204. However their accretion rates
 are comparable. Although \citet{2012MNRAS.426..851K} suggested that larger mass
 BHs are associated with smaller accretion rates, this does not persist for
 all types of quasars (e.g., J2335$-$0927, J1623$+$3419, J1433$+$3209;
 \citealt{2012MNRAS.426..851K}). Hence it appears that the BH mass and accretion
 rates of these DDRQs are quite similar to other quasars.

 The discovery of a new DDRQ underlines the importance of identifying more such
 objects to study the properties of these objects at different wavelengths to
 estimate time scales of episodic activity, and properties of host galaxies and
 their environments to understand the triggering mechanisms for jet activity.

\acknowledgments  
 We are thankful to Hum Chand, Jean Surdej and Jacopo Fritz for
 several useful discussions. We thank also the anonymous referee
 for her/his valuable comments. SN and RR are funded by the BELSPO grant to
 conduct their scientific researches. HCC is thankful to the UGC (New Delhi) 
 for the research project financial assistance.
 The GMRT is a national facility operated by the National Centre for
 Radio Astrophysics of the Tata Institute of Fundamental Research. 
 The National Radio Astronomy Observatory is a facility of the National 
 Science Foundation operated under co-operative agreement by Associated 
 Universities Inc. This research has made use of the NASA/IPAC extragalactic 
 database (NED) which is operated by the Jet Propulsion Laboratory, Caltech, 
 under contract with the National Aeronautics and Space Administration.
 Funding for the SDSS and SDSS-II has been provided by the Alfred P. Sloan
 Foundation, the Participating Institutions, the National Science Foundation,
 the U.S. Department of Energy, the National Aeronautics and Space
 Administration, the Japanese Monbukagakusho, the Max Planck Society, and the
 Higher Education Funding Council for England.\\  

% Bibliography ................... 
%___________________________________________________________________________

%______________________________
\clearpage
%_________________________________________________________________________________

\end{document}